\documentclass[aip,cha,amsmath,amssymb,reprint]{revtex4-1}

\usepackage{graphicx,color}

\definecolor{red}{rgb}{1,0,0}
\definecolor{green}{rgb}{0,1,0}
\definecolor{blue}{rgb}{0,0,1}

\begin{document}

\title{Stochastic modelling of non-stationary financial assets}

\author{Joana Estevens}
\affiliation{Departamento de Matem\'atica
  and Centro de Matem\'atica e Aplica\c{c}\~oes Fundamentais,
  Faculdade de Ci\^encias, 
  University of Lisbon, Campo Grande,
  1749-016 Lisboa, Portugal}
\author{Paulo Rocha}
\affiliation{Departamento de Matem\'atica
  and Centro de Matem\'atica e Aplica\c{c}\~oes Fundamentais,
  Faculdade de Ci\^encias, 
  University of Lisbon, Campo Grande,
  1749-016 Lisboa, Portugal}
\author{Jo\~ao P.~Boto}
\affiliation{Departamento de Matem\'atica
  and Centro de Matem\'atica e Aplica\c{c}\~oes Fundamentais,
  Faculdade de Ci\^encias, 
  University of Lisbon, Campo Grande,
  1749-016 Lisboa, Portugal}
\author{Pedro G.~Lind}
\affiliation{Institut f\"ur Physik, Universit\"at Osnabr\"uck,
             Barbarastrasse 7, 49076 Osnabr\"uck, Germany}

\date{\today}

\begin{abstract}
We model non-stationary volume-price distributions with a log-normal 
distribution and collect the time series of its two parameters.
The time series of the two parameters are shown to be stationary and 
Markov-like and consequently can be modelled with Langevin equations, 
which are derived directly from their series of values.
Having the evolution equations of the log-normal parameters, we 
reconstruct the statistics of the first moments of volume-price 
distributions which fit well the empirical data.
Finally, the proposed framework is general enough to study other
non-stationary stochastic variables in other research fields,
namely biology, medicine and geology.
\end{abstract}



\maketitle

\begin{quotation}
While several methodologies are available for modelling
stationary processes, nature and complex processes in
nowadays society is typically non-stationary.
Examples range from the most fundamental turbulent fluids,
weather dynamics and wind energy systems to human mobility,
the brain signals and finance.
In this paper we attack the problem of modelling non-stationary
stochastic variables, addressing the specific example of financial
volume-prices.
We introduce a framework to model the evolution of non-stationary time
series, applying it to the specific case of the volume-price of
financial assets from 1750 companies at the New York Stock Exchange,
collected from Yahoo database every 10 minutes during more than two years.
Since the volume-price is typically a non-stationary stochastic variable
but follows the same functional distribution, we assume that all time
dependence of the variable is incorporated in time fluctuations of
the distribution parameters, which are themselves stationary and
evolve as coupled variables.
\end{quotation}

\section{Introduction}
\label{intro}

When addressing stochastic processes in nature, one of the fundamental
assumptions is its stationary or non-stationary character.
If the process is stationary, there is a probability density function
associated to it, that does not change in time.
In this case, the laws underlying series of measurements from the process
can be assessed through proper averaging and computations of value series
within finite size time-windows.
However, the typical case found in nature is to observe processes
whose probability density function changes also with time, making the
derivation of the underlying laws more difficult.

In this paper we address a concrete example of a non-stationary
stochastic process and show how to reproduce its statistical
features, assuming not the strong condition of having time-invariant
probability densities but only that all associated probability densities
have a fixed functional form. In other words, having defined the function
that best fits the probability density functions, its parameters are
function of time. In particular, we will show that the parameters
vary stochastically in time and follow a stationary process, which enables
one to recover the non-stationary statistical moments characterizing the
original process. 

\begin{figure*}[t]
\centering
\includegraphics[width=0.95\textwidth]{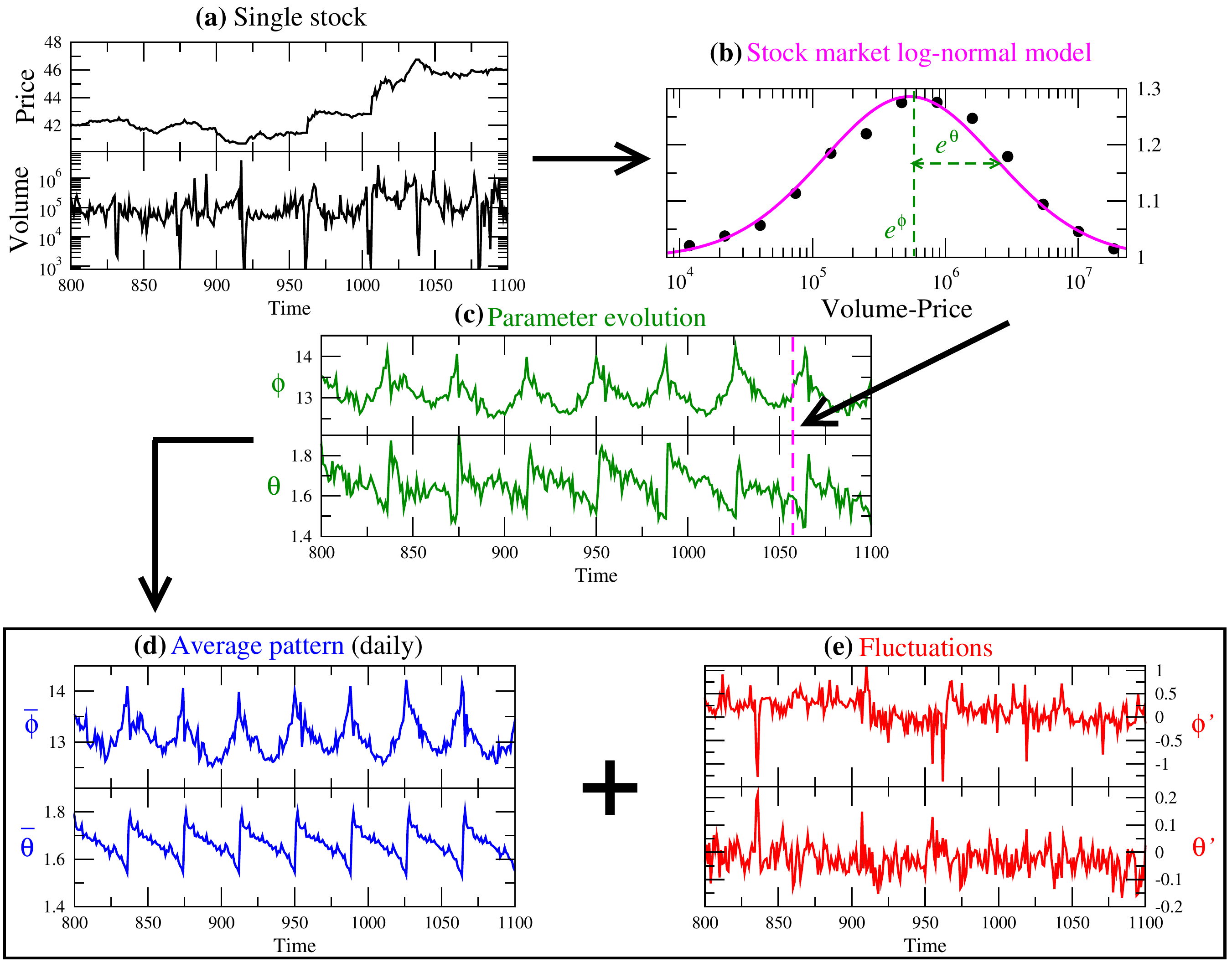}
\caption{\protect  
         We start with the price and volume series of 1750 companies.
         In {\bf (a)} we can see the volume and price series for 
         just one of the companies. Multiplying the two series 
         yields the volume-price series which follows a log-normal 
         distribution with parameters $\phi$  and $\theta$. 
         In {\bf (b)} we can see the empirical density represented 
         by the dots and the adjusted log-normal to the data, solid 
         line, for a particular window of 10 minutes. 
         Each 10-minutes window yields a log-normal with different 
         parameters. 
         Thus, we will have {\bf (c)} a time series for the parameter 
         $\phi$ and another one for $\theta$. 
         Each time series can be decomposed in 
         {\bf (d)} a daily averaged pattern and 
         {\bf (e)} fluctuations around this pattern. We will describe 
         the evolution observed in (c) by analyzing this components 
         separately.}
\label{fig01}
\end{figure*}

The framework here introduced is related to the so-called superstatistics
approach introduce by Beck in the early nineties \cite{superstatistics}.
Superstatistics is a branch of statistics aimed to the study of non-linear
and non-equilibrium systems. Complex systems often show a behavior which
can be regarded as a superposition of different
dynamics\cite{superstatistics}.
We study the specific case of volume-price distributions
in the New York Stock Exchange (NYSE), collected directly 
from Yahoo Finance.
While the asset price shows how valuable the asset is and
volume accounts for the corresponding market liquidity for that
asset, the volume-price incorporates the interaction between both
financial quantities, retrieving the total capitalization being
transitioned in the market. 
Figure \ref{fig01} shows the overview of the framework here proposed,
with Fig.~\ref{fig01}a illustrating the time-series of the price and
volume of one single company.

We start in Sec.~\ref{sec:model} by assuming the log-normal 
distribution as the model for fitting each volume-price distribution 
from Yahoos' database.
Such assumptions is in line with previous findings\cite{Paulo}.
An example of the log-normal volume-price distribution is shown
in Fig.~\ref{fig01}b.
The dots represent the empirical
probability density function of the logarithm of the volume-price
time series and the solid line is the theoretical log-normal
distribution that best fits the data.
Since the volume-price is non-stationary, the volume-price
distribution varies in time, i.e.~the two parameters defining the
log-normal fit are time dependent, as illustrated in Fig.~\ref{fig01}c.
We then study the evolution of both parameters, considering that the 
two parameters include {\it all} the time dependency of the
volume-price variable, and decompose the evolution
of each parameter, into two separated additive contributions:
an average behavior (Fig.~ \ref{fig01}d) and the corresponding
stochastic fluctuations around it (Fig.~ \ref{fig01}e).
While the average behavior is modelled with a polynomial in
time for the time span of one day, the fluctuations are modelled
through a system of two coupled stochastic differential 
equations\cite{Friedrich}.
In particular, 
by knowing how the fluctuations of both parameters evolve in time,
we will be able to retrieve the non-stationary evolution of the
volume-price in the NYSE, namely the evolution of its statistical
moments.
Finally, in Sec.\ref{sec:conclusions} we summarize the main
conclusions and discuss possible applications of this work to
other fields of study.

\section{Log-normal model for volume-prices}
\label{sec:model}

In the following we analyze the volume-price series of 
1750 companies having listed shares in the New York Stock Exchange
(NYSE), with a sampling frequency of $0.1$ min$^{-1}$.
After removing all the after-hours trading and discarding all the days 
with recorded errors, our dataset contains 17708 data points for each 
company covering a total period of $976$ days.
Figure \ref{fig01}a shows an illustrative example of the price and
volume series of one specific company.
All the data were collected from the website 
{\tt http://finance.yahoo.com/} and more details concerning its
preprocessing may be found in the references of Rocha and 
co-workers\cite{paulo1,paulo2,Paulo}. 
\begin{figure}[t]
\centering
\includegraphics[width=0.45\textwidth]{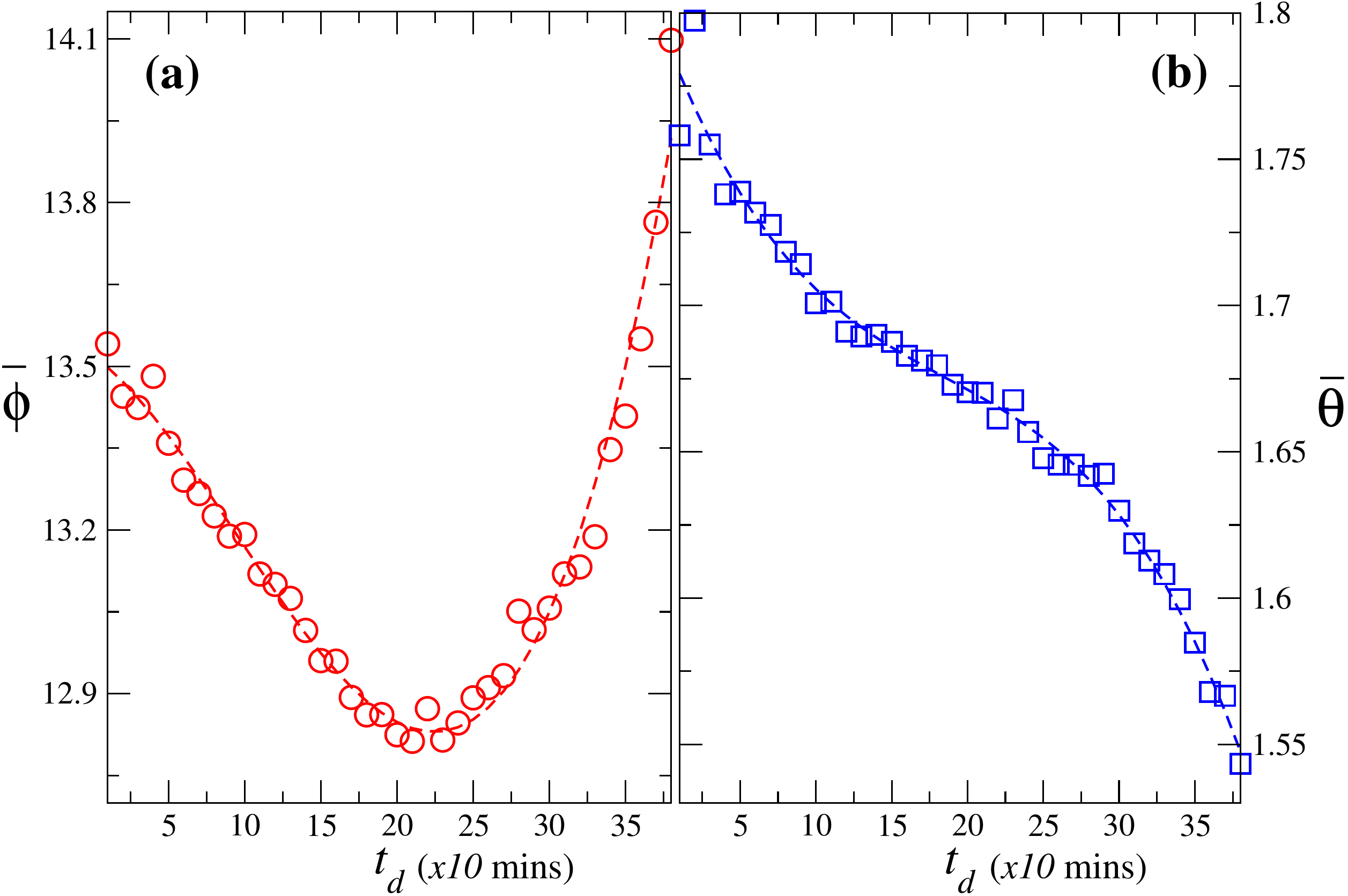}
\caption{\protect  
         The average over all trading days of 
         {\bf (a)} parameter $\phi$ and 
         {\bf (b)} parameter $\theta$, with the corresponding 
         fitting cubic polynomials (see text).}
\label{fig02}
\end{figure}

In previous works, it was shown that the log-normal distribution 
properly models the distribution of volume-price in each $10$-minute
frame\cite{Paulo}. 
The  probability density function (PDF) of the log-normal is:
\begin{equation}
p_{\phi,\theta}(s)= \frac{1}{s \sqrt{2\pi}\theta}
               \exp{\left[-\frac{(\log s-\phi)^2}{2\theta^2}\right]} 
\label{Log-normal_PDF}
\end{equation}
and is one of the common models in finance data\cite{Paulo,silvio}.
Here $s$ symbolizes the volume-price.
Parameter $\phi$ accounts for the mean and  $\theta$ for the standard 
deviation of the logarithm volume-price series and they take
different values for each $10$-minute frame.
Consequently, fitting the volume-price distributions with 
Eq.~(\ref{Log-normal_PDF}) at each $10$-minute lag for the full 
$976$-day period yields two data series of values, one for
$\phi$ and another for $\theta$.

\subsection{Average behavior of the volume-price distribution}

The original time series of $\phi$ and $\theta$ parameters
is decomposed
into the sum of their average daily patterns, $\overline{\phi}$ and 
$\overline{\theta}$, with the respective fluctuations, $\phi^\prime$ and 
$\theta^\prime$, around the average patterns:
\begin{subequations}
\begin{eqnarray}
\phi &=& \overline{\phi}+ \phi^{\prime} \, , \label{phidec}\\   
\theta &=& \overline{\theta}+ \theta^{\prime} \label{thetadec}  \, .
\end{eqnarray}
\end{subequations}

In Figure \ref{fig02} one sees the average patterns of $\phi$ and 
$\theta$, from which one can extract insight concerning the 
typical behavior of trades during the opening time, as follows.
In Fig.~\ref{fig02}a the average pattern of parameter 
$\phi$ is plotted. One notices that trading is heavier, i.e.~larger 
values of $\phi$, at the beginning and end of the day compared to
the rest of the day.
Such feature reflects the fact that the opening and the closing of the 
NYSE are very peculiar times.
In the beginning of the day, the volume-price series has high values,
because it concentrates the information of the traders during the 
precedent closing period when traders postpone their next buy or sell
to the next opening moment.
After the opening one observes an approximately linear decrease of 
the trading activity, which in the second half of the day increases
cubically.
This higher rate in the end of the day may happen because traders 
react against a deadline, the closing time.

Close to the end of the day, volume-prices start to grow again, since
traders try their last chance to buy and sell according to the market 
present state in a sort of herd behavior.
This herd behavior can also be derived from the daily pattern of
parameter $\theta$ plotted in Fig.~\ref{fig02}b.
In the beginning of the day there is a large variance in our data
(large $\theta$), due to the different perceptions the traders
have following a closing period where decisions were taken 
following different strategies and
based in different alternative sources of information (e.g.~open
exchange markets elsewhere in the world).
After that it decreases monotonically reflecting the tendency
of traders to behave in the same way (small volume-price variance),
since they based their decisions in the same real time prices of the
NYSE, attaining a minimum at the closing time.
Furthermore, during the day the standard deviation relaxes, showing
a more constant value around noon. This value of $\theta$ in the middle
of a (typical) day, being less sensitive to the opening and closing
moments, defined through an inflexion point,
should be characteristic of the specific stock exchange we
are analyzing.

Both average patterns are well approximated, in mean-square sense, 
by cubic polynomials of time:
\begin{subequations}
\begin{eqnarray}
\bar{\phi}(t_d) &=& a_{\phi}t_d^3+b_{\phi}t_d^2+c_{\phi}t_d+d_{\phi} 
\label{phibar} \,, \\
\bar{\theta}(t_d) &=& a_{\theta}t_d^3+b_{\theta}t_d^2+c_{\theta}t_d+d_{\theta} 
\label{thetabar} \,,
\end{eqnarray}
\label{xTheta}%
\end{subequations}
where $t_d= (t \pmod{144})-54$ in units of $u=10$ minutes, and
$a_{\phi}=8.2\times 10^{-5}$,
$b_{\phi}=-2.3\times 10^{-3}$,
$c_{\phi}=-2.0\times 10^{-2}$,
$d_{\phi}=13.52$, 
$a_{\theta}=-1.0\times 10^{-5}$,
$b_{\theta}=5.6\times 10^{-4}$,
$c_{\theta}=-1.3\times 10^{-2}$ and
$d_{\theta}=1.79$.

Note that the market is only open for normal trading during 
$6$h$30$min ($39\times 10$min).
Outside of the normal trading period we consider
$\bar\phi$ and $\bar\theta$ to be zero.
These cubic models for the average daily pattern, depend on the data set
being analyzed, i.e.~on the stock market, and typical maxima and 
minima values of both parameters, as well as their inflexion points can be
straightforwardly estimated.

\subsection{Modelling stochastic fluctuations of the parameters}

The fluctuations around the patterns in Fig.~\ref{fig02} 
were obtained by subtracting the 20-day moving average pattern
(Fig.~\ref{fig01}d) from the corresponding parameter
(Fig.~\ref{fig01}c), yielding the fluctuations
$\phi^\prime$ and $\theta ^\prime$ illustrated in Fig.~\ref{fig01}e.

\begin{figure}[t]
\centering
\includegraphics[width=0.49\textwidth]{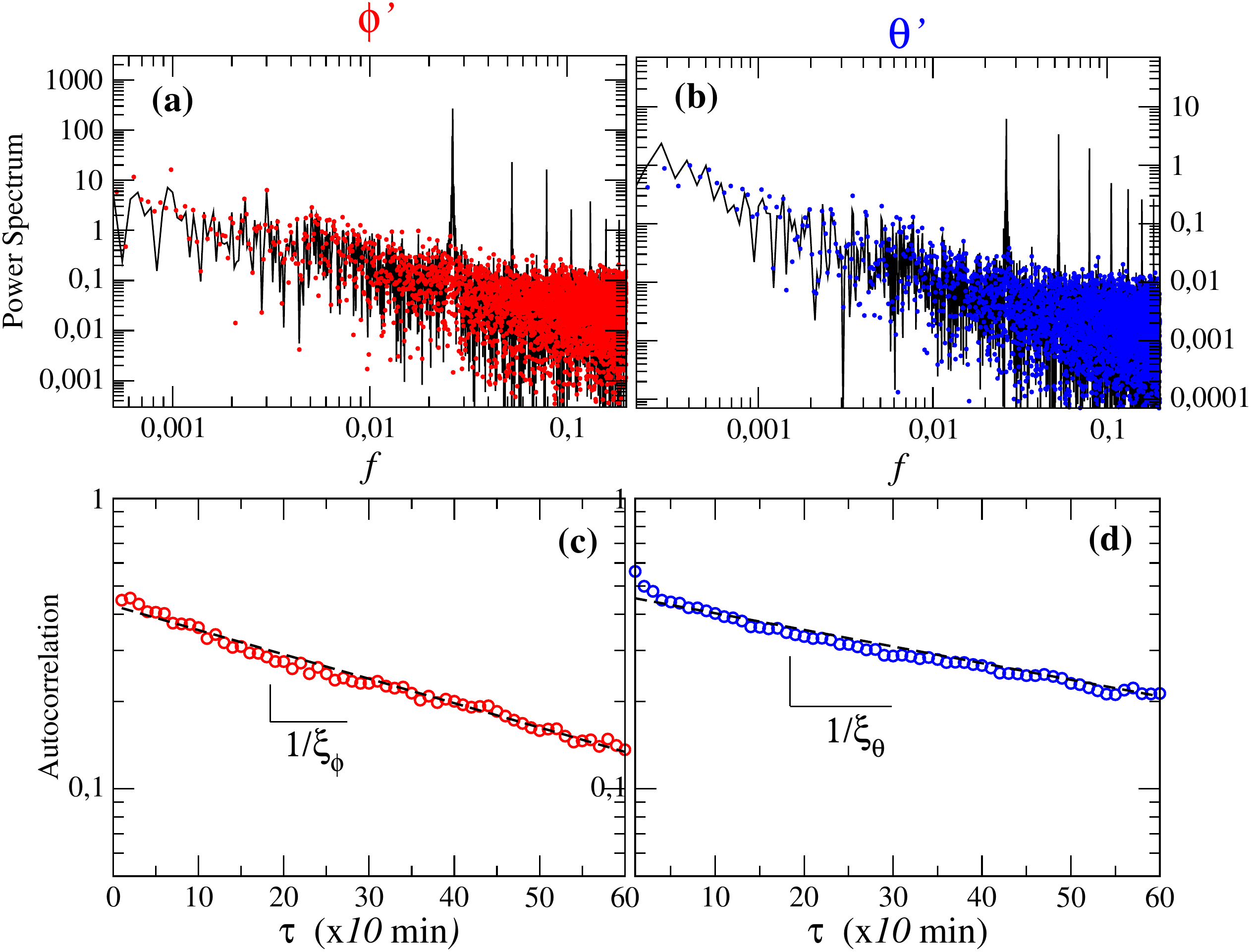}
\caption{\protect
         {\bf (a)} Power spectrum of the $\phi$ time series and 
         {\bf (b)} of the $\theta$ series, both detrended (see text). 
         {\bf (c)} Autocorrelation function (circles) and linear 
                   function fitted to the data (dashed lines) in a 
                   log-lin scale for the $\phi$ time series and 
                   {\bf (d)} for the $\theta$ time series.}
\label{fig03}
\end{figure}

Important for the modelling of the fluctuations is to separate them
from all periodic modes of the time variation of each parameter.
Figures \ref{fig03}a and \ref{fig03}b show the power spectrum
of the original parameters (black lines) and the corresponding
fluctuations (colored points). As one clearly sees, the periodic 
peaks, observed for the original parameter, are not present in the
spectra of their fluctuations, showing that the periodic behavior
is properly detrended.

In Figs.~\ref{fig03}c and \ref{fig03}d we plot the autocorrelation 
function $\alpha$ for the fluctuations $\phi^\prime$ and $\theta^\prime$ 
respectively, showing a clear exponential decay which is modelled as
\begin{equation}
\label{ACFeq}
\alpha_{\phi^{\prime}, (\theta^{\prime})}= \beta_{\phi^{\prime}, (\theta^{\prime})} 
         \exp{\left ( -\frac{\tau}{\xi_{\phi^{\prime},(\theta^{\prime})}} \right )} .
\end{equation}%
For $\phi^{\prime}$ one obtains $\xi_{\phi^{\prime}}=52.08\times 10$ mins
and $\log(\beta_{\phi^{\prime}})= -0.8496$, while for $\theta^\prime$ one obtains 
$\xi_{\theta^{\prime}}=75.76\times 10$ s and $\log(\beta_{\theta})= -0.7776$. 
In other words, one surprisingly finds that,
while the logarithm average fluctuation $\phi^{\prime}$
looses memory beyond one daily cycle of the stock market (open during
9 hours) the fluctuation of lognormal variances shows a
memory beyond the closing of the stock market, probably incorporating
information from after-hour tradings.
\begin{figure}[t]
\centering
\includegraphics[width=0.45\textwidth]{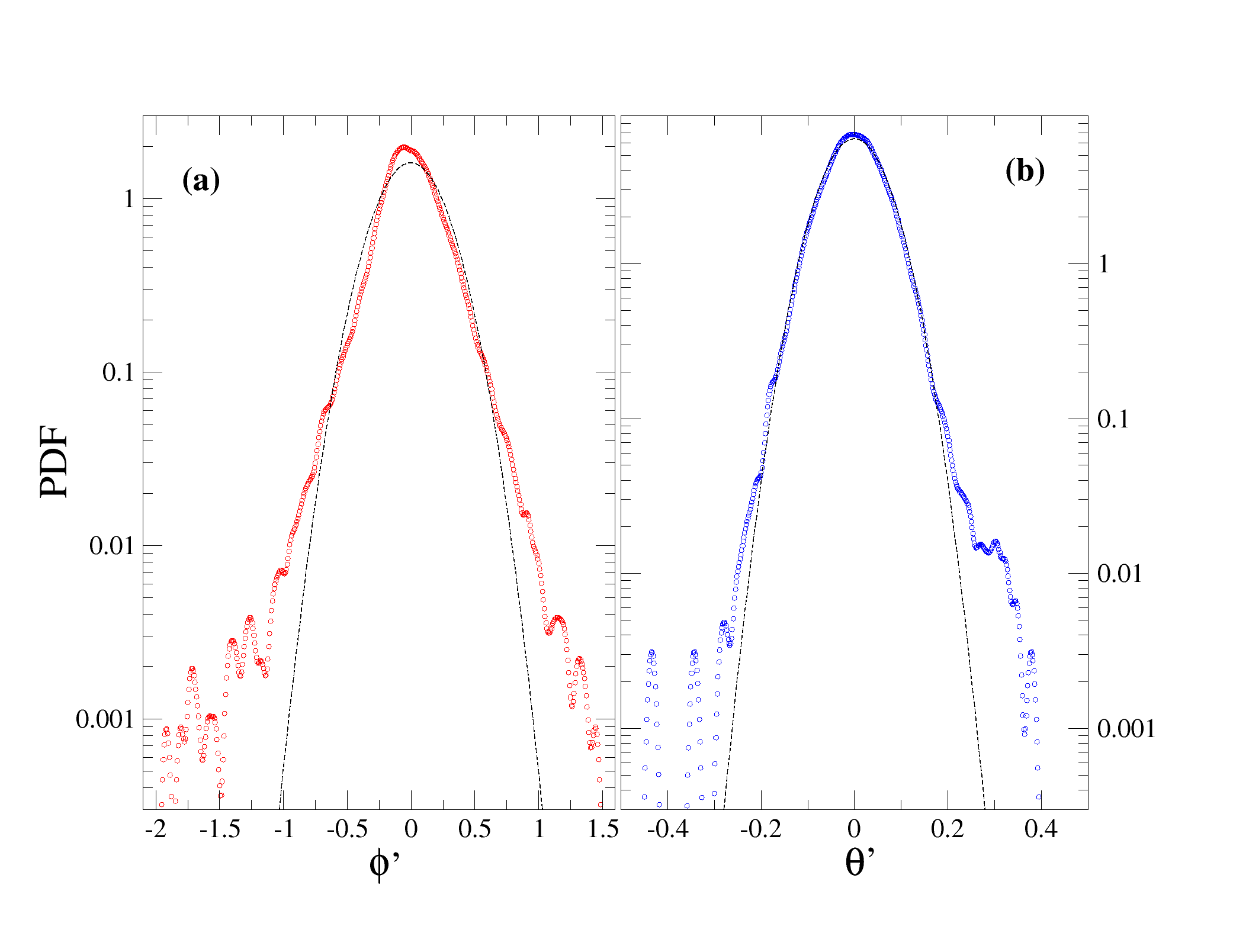}
\caption{\protect
         Marginal probability density function (black) and Gaussian 
         adjusted PDF (red) of the detrended 
         {\bf (a)} $\phi$ parameter and 
         {\bf (b)} $\theta$ parameter time series.}
\label{fig04}
\end{figure}
\begin{figure}[t]
\centering
\includegraphics[width=0.24\textwidth]{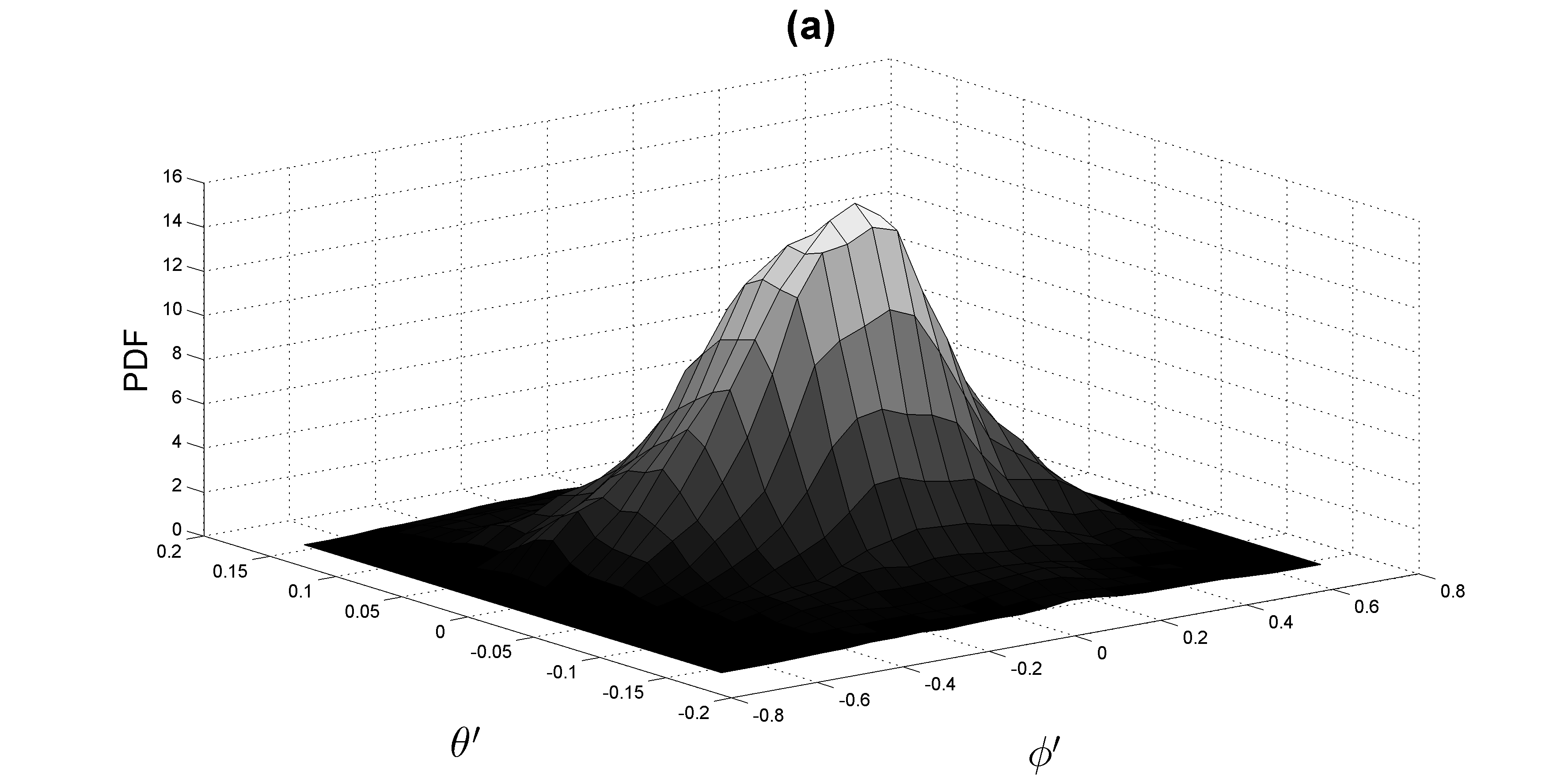}%
\includegraphics[width=0.24\textwidth]{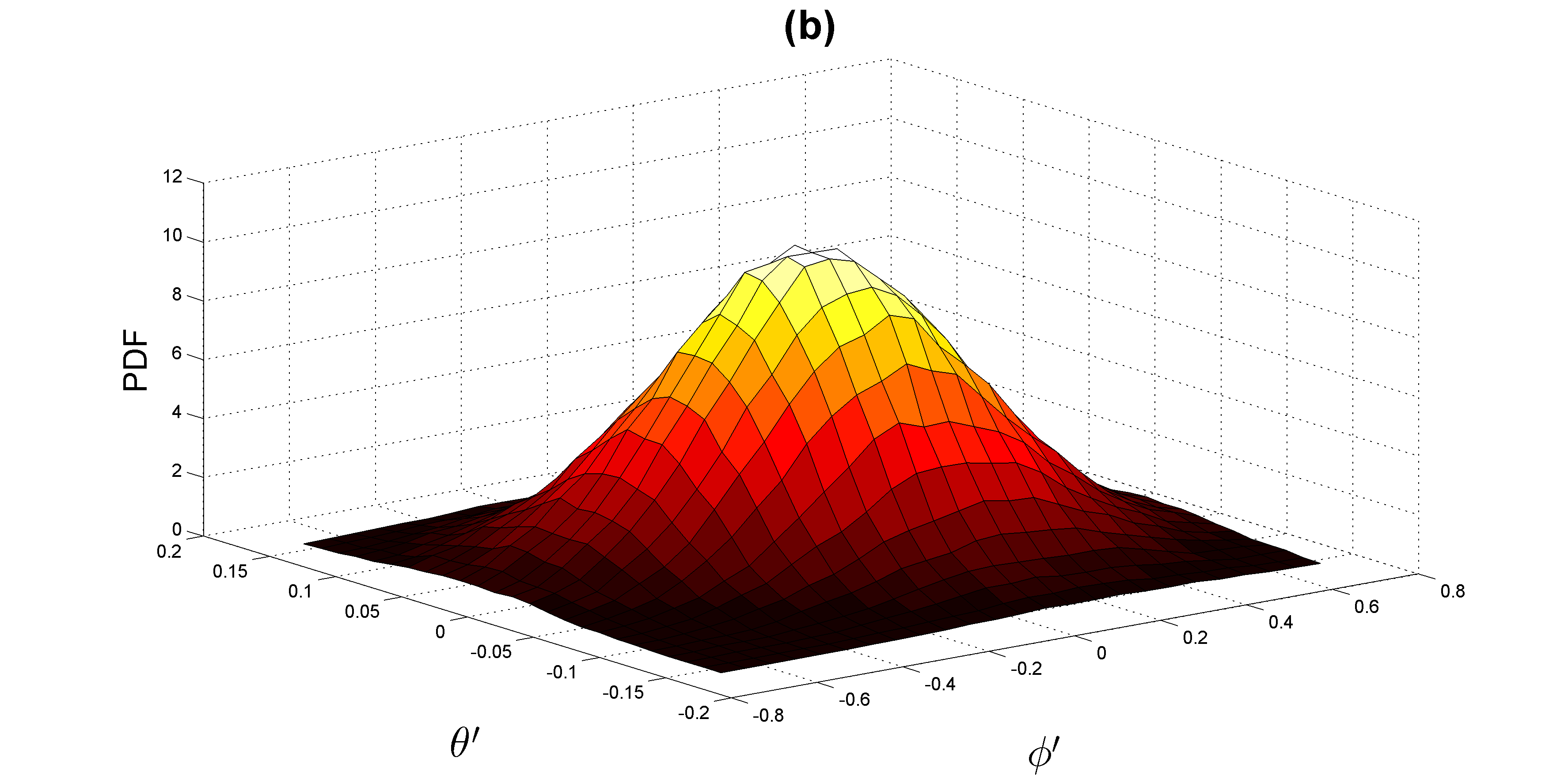}
\includegraphics[width=0.48\textwidth]{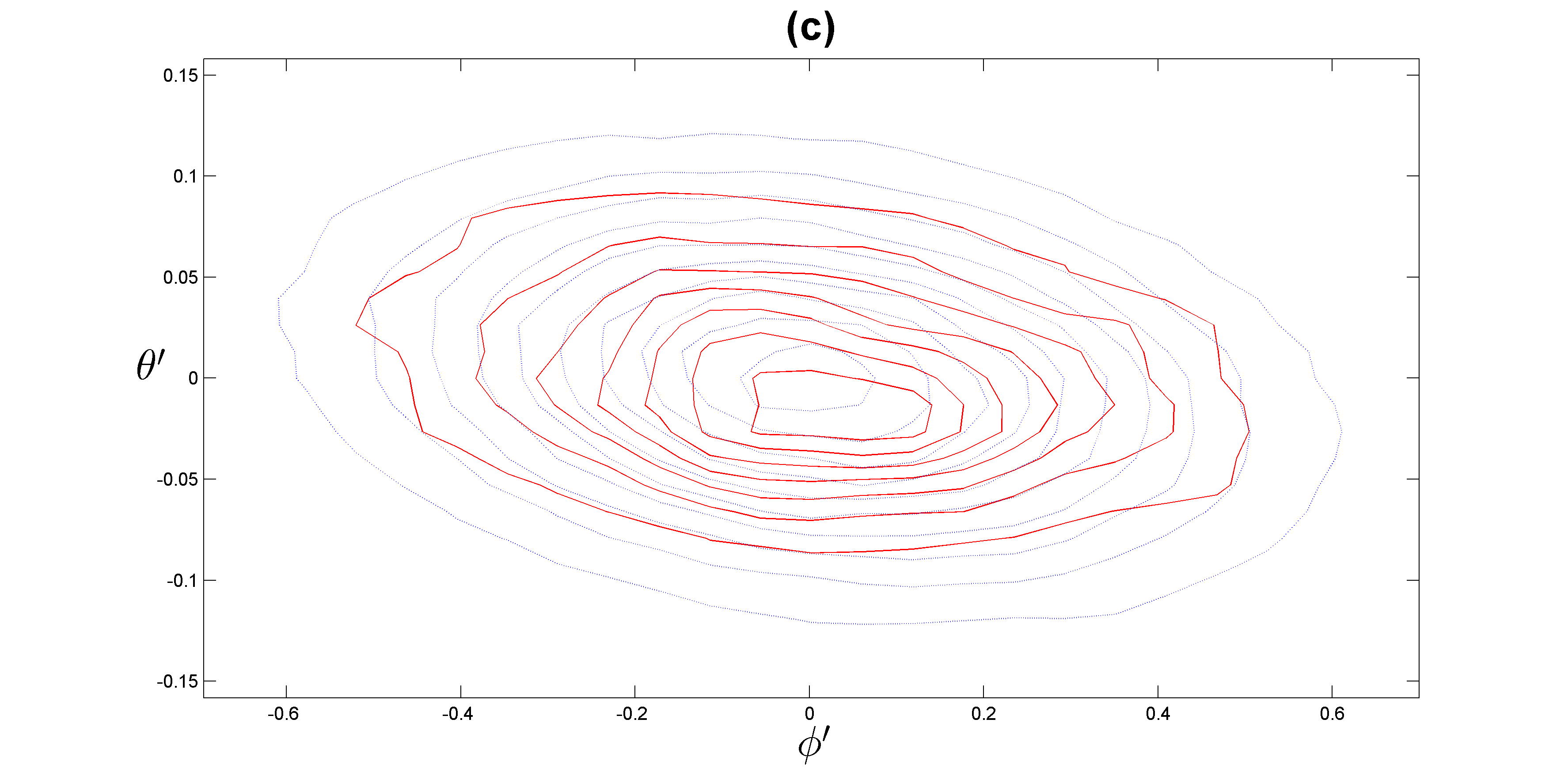}
\caption{\protect
         \textbf{(a)} Joint PDF of the empirical time series and
         \textbf{(b)} a multivariate normal distribution with 
         mean vector and covariance matrix equal to the ones 
         of our empirical data. \textbf{(c)} Contour plot of both
         empirical and theoretical joint PDFs in (a) and (b)
         respectively.}
\label{fig05}
\end{figure}
\begin{figure*}[htb]
\centering
\includegraphics[width=0.32\textwidth]{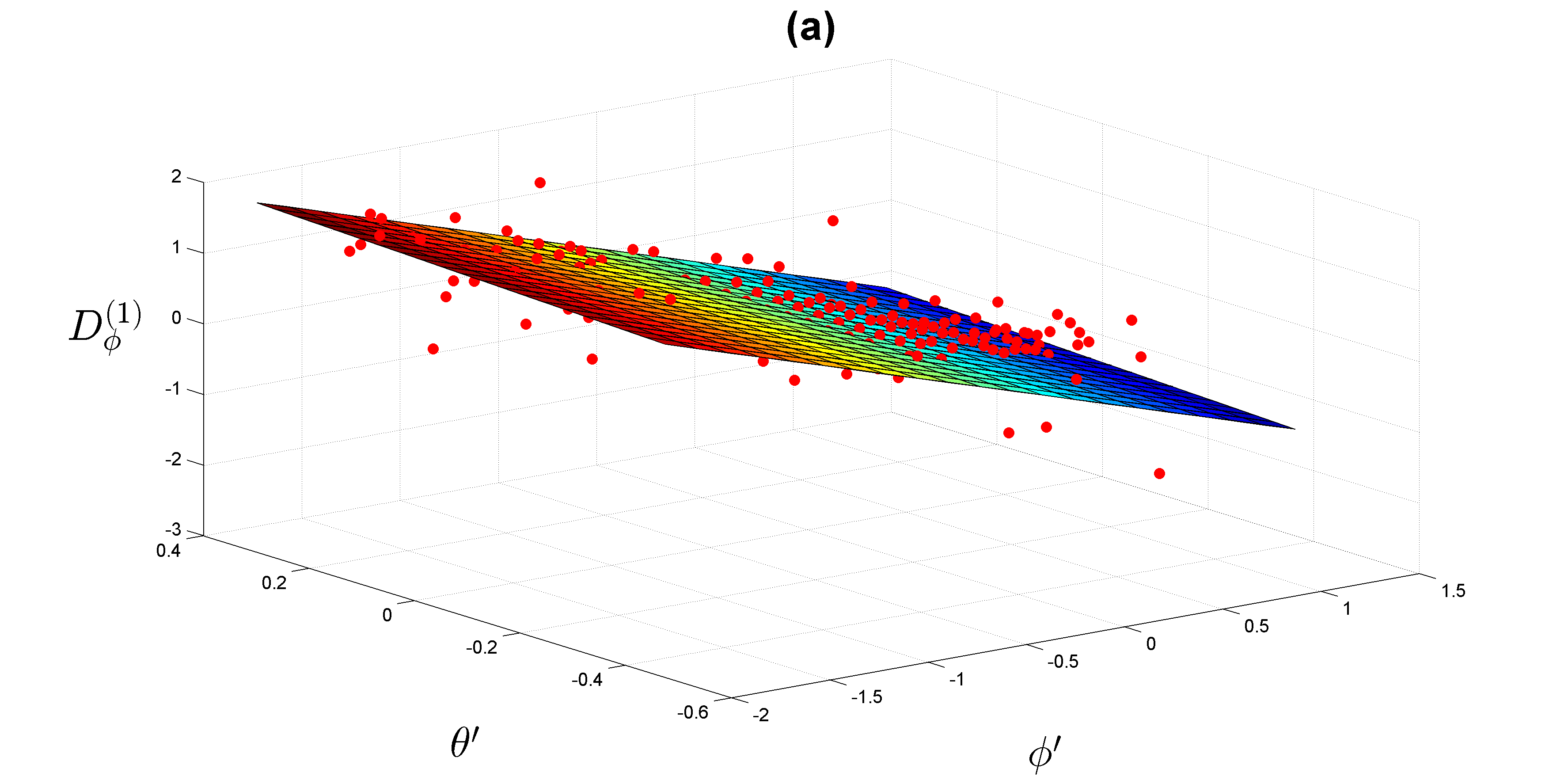}%
\includegraphics[width=0.32\textwidth]{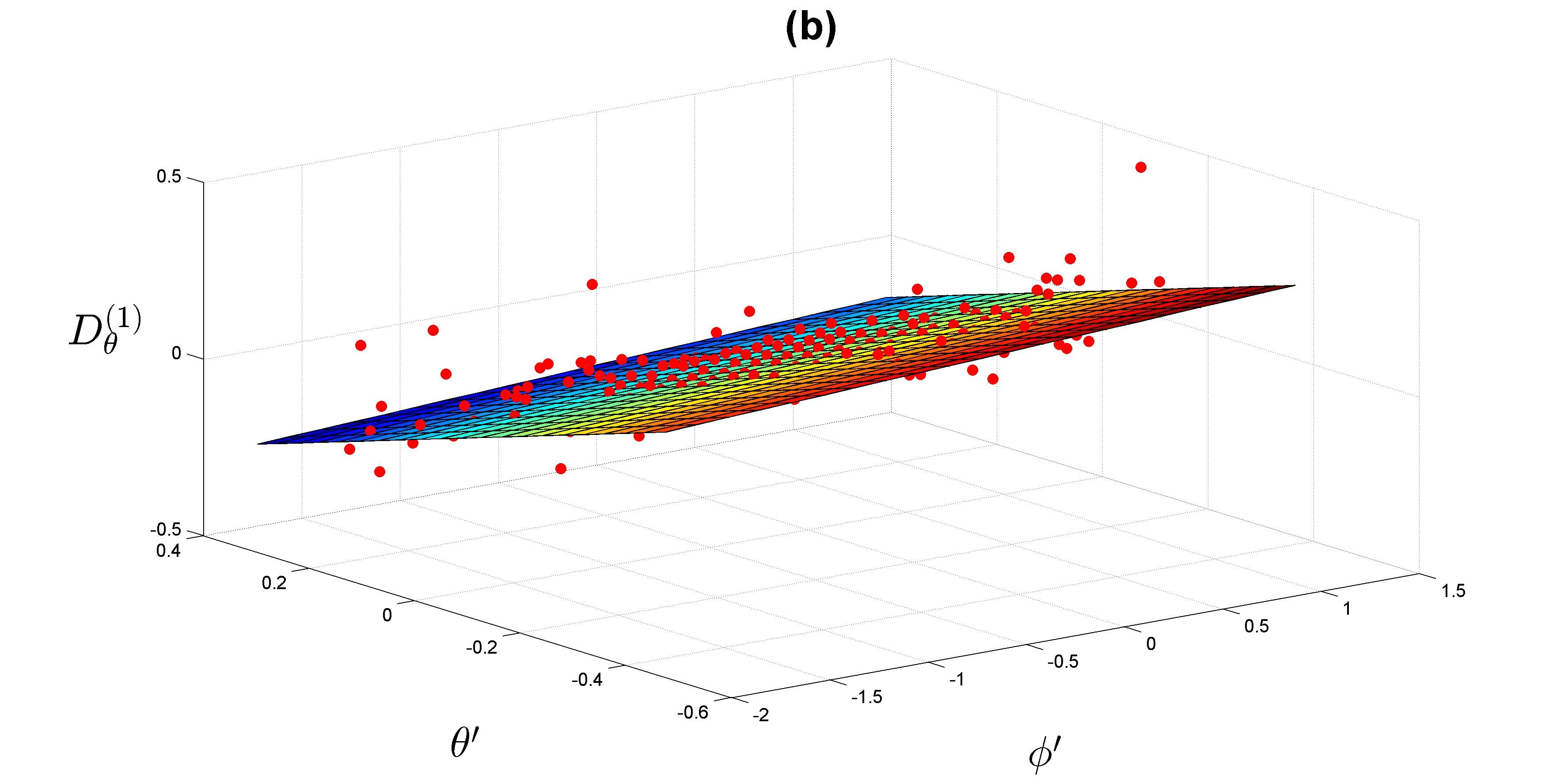}
\includegraphics[width=0.32\textwidth]{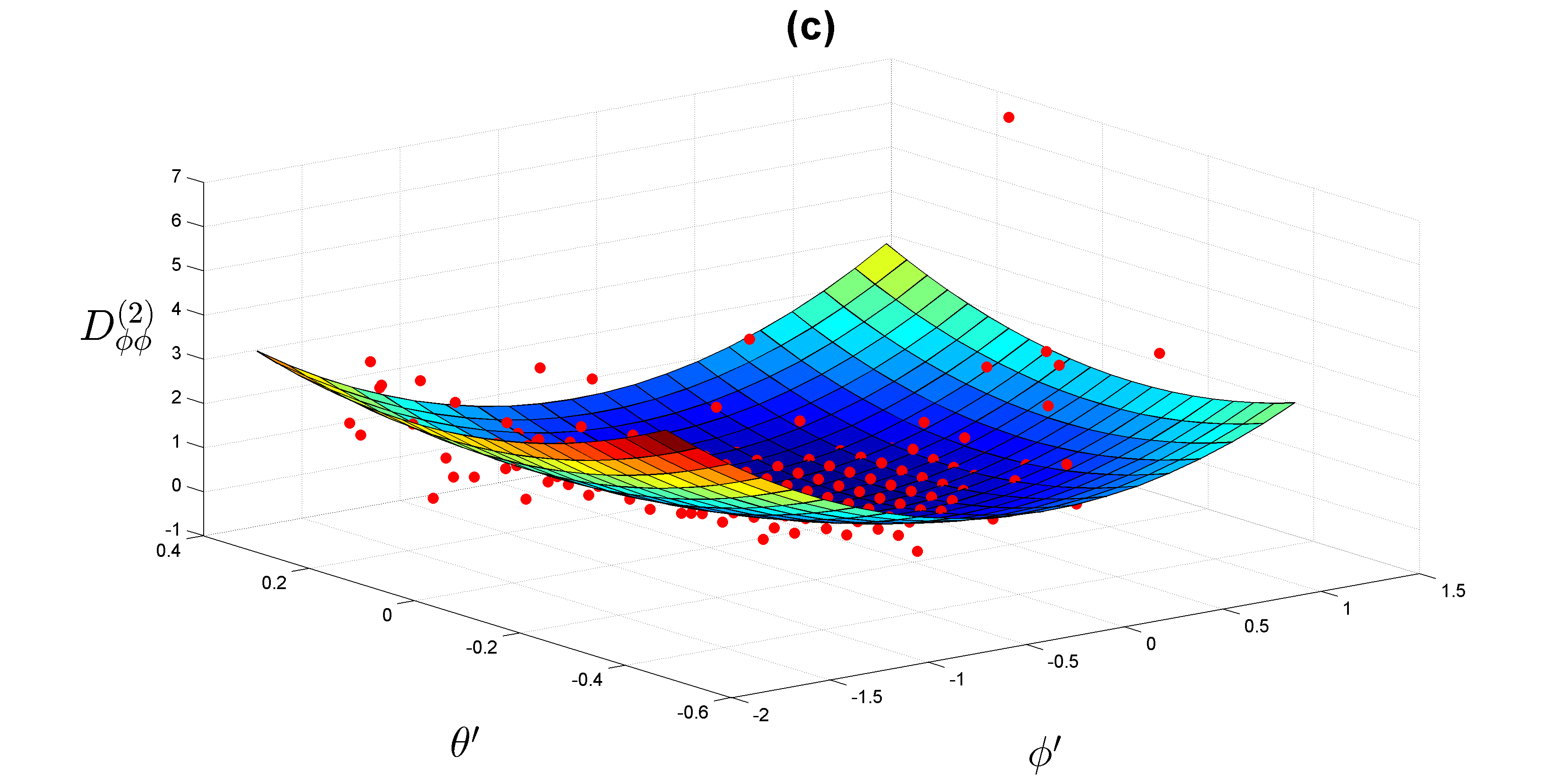}%
\includegraphics[width=0.32\textwidth]{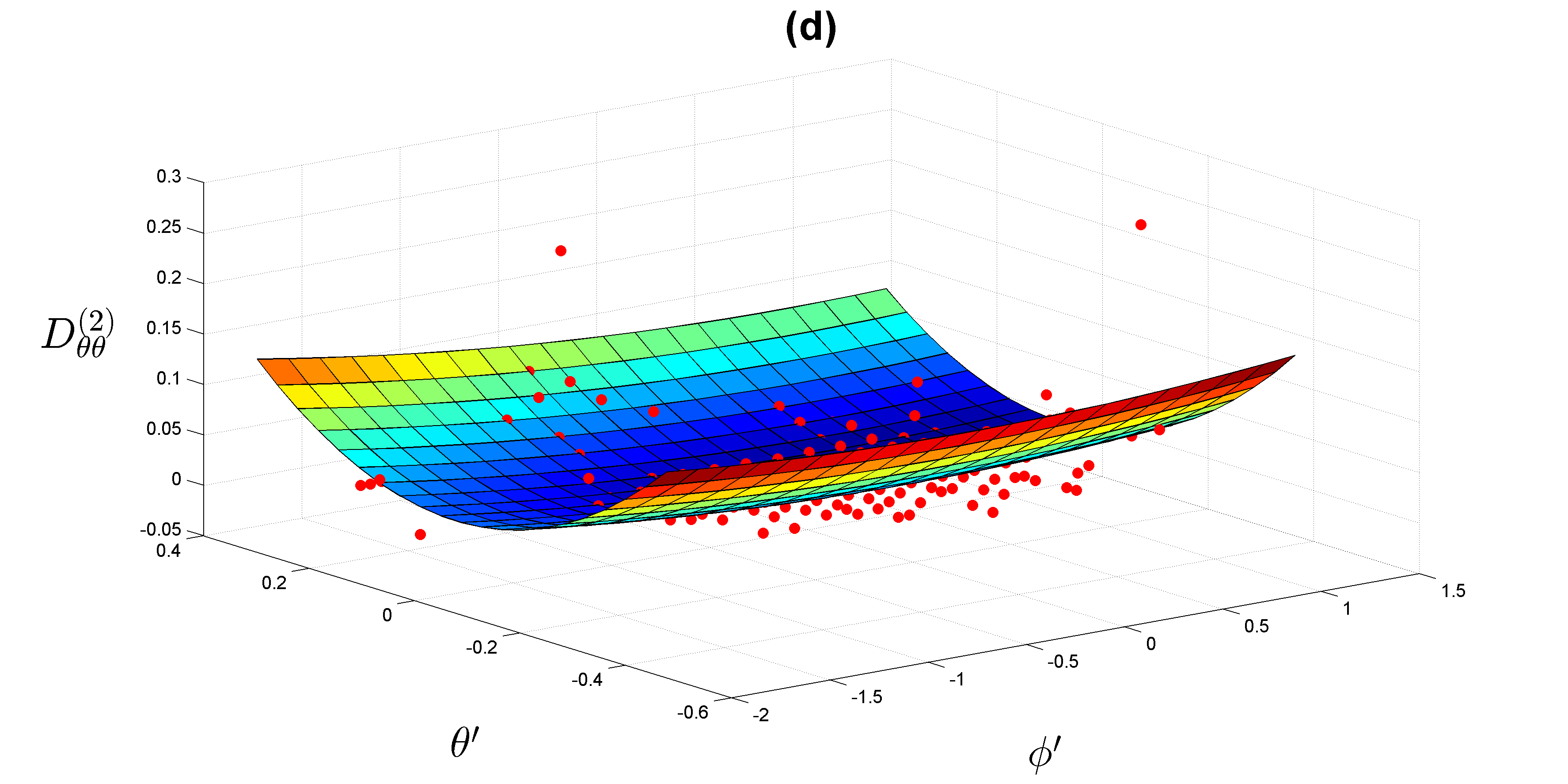}%
\includegraphics[width=0.32\textwidth]{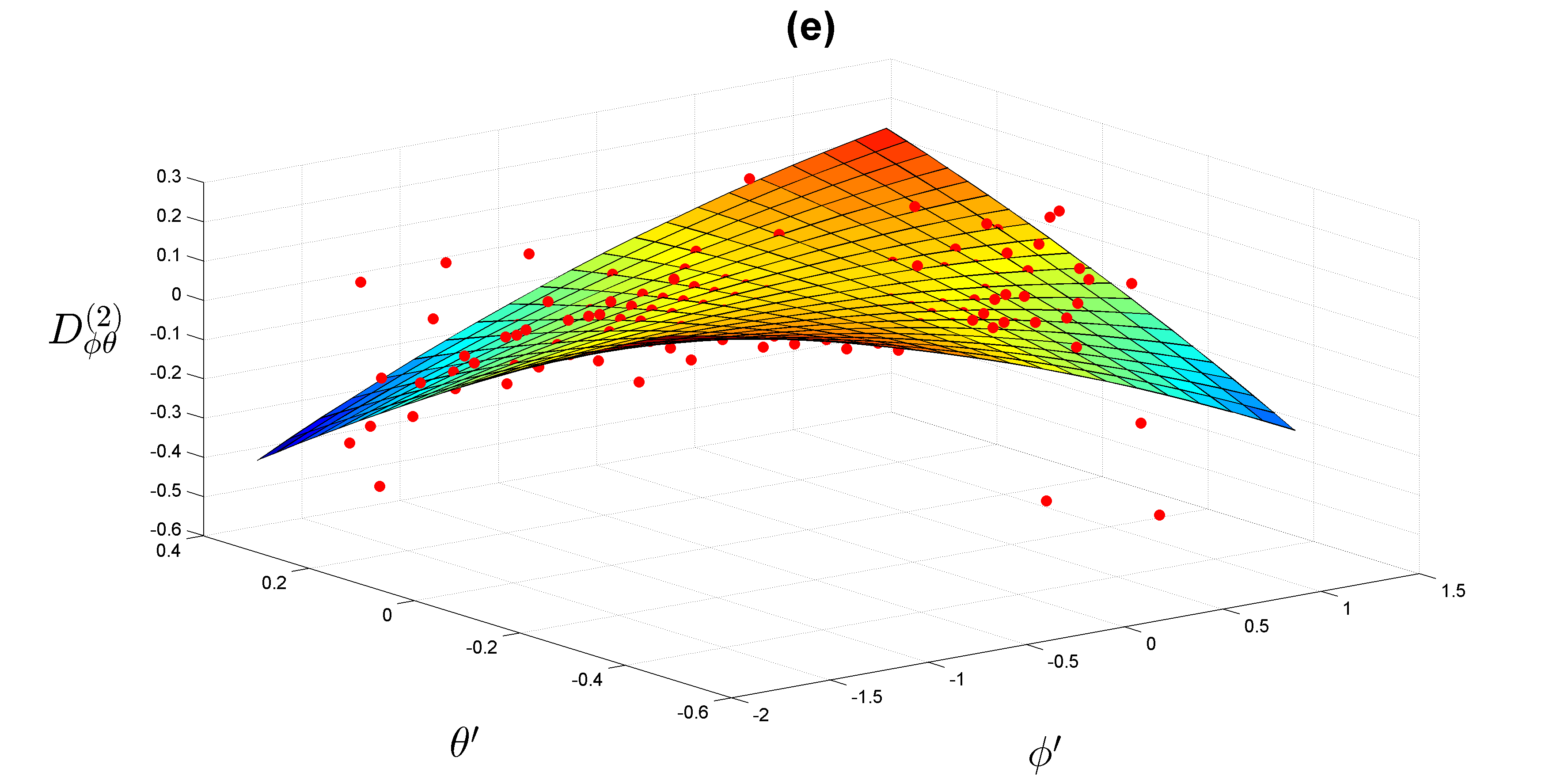}
\caption{\protect
         {\bf (a)} Drift coefficient of the $\phi$ component, ${D_\phi}^{(1)}$,
         {\bf (b)} Drift coefficient of the $\theta$ component, 
                   ${D_\theta}^{(1)}$.
         The three components of the diffusion matrix: 
         {\bf (c)} function $D^{(2)}_{\phi \phi}$, 
         {\bf (d)} function $D^{(2)}_{\theta \theta}$ and 
         {\bf (e)} function $D^{(2)}_{\phi \theta}$.}
\label{fig06}
\end{figure*}

To end this section we show that both fluctuations, $\phi^{\prime}$
and  $\theta^{\prime}$, are stochastic variables with a joint Gaussian
distribution and anti-correlated with each other.

Figures \ref{fig04}a and \ref{fig04}b show the marginal PDF
of the fluctuations $\phi^\prime$ and $\theta^\prime$ respectively
(colored symbols) compared with the respective Gaussian, having the same
mean and variance (dashed lines).
The central region of the marginal PDF of $\theta^{\prime}$ is in good
agreement with the Gaussian function, presenting deviations at the
most outer regions. 
Apparently, for $\phi^\prime$ the Gaussian approximation is not as good as for
$\theta^{\prime}$. However, the deviations from the Gaussian functional
shape of the marginal distribution of $\theta^{\prime}$ is due to the fact
that both parameter fluctuations are correlated as can be seen in
Fig.~\ref{fig05}.

Indeed, computing the covariance matrix
\begin{equation}
  \mathbf{\Sigma} =
\left [
  \begin{array}{cc}
    \Sigma_{\phi^{\prime}\phi^{\prime}} &   \Sigma_{\phi^{\prime}\theta^{\prime}} \cr
  \Sigma_{\theta^{\prime}\phi^{\prime}} &   \Sigma_{\theta^{\prime}\theta^{\prime}} \nonumber
  \end{array}
\right ] =
\left [
  \begin{array}{cc}
  0.0619 & -0.0036 \cr
  -0.0036 &  0.0039 \nonumber
  \end{array}
\right ] .
\label{sigma}
\end{equation}
Being a non-diagonal matrix, the covariance matrix shows that both
parameter fluctuations are correlated.
Consequently, the joint distribution $\rho (\phi^{\prime},\theta^{\prime})$
of both correlated parameters can be well fitted with a two-dimensional
Gaussian distribution of correlated variables as\cite{statisticsbook}:
\begin{equation}
  \rho(\phi^{\prime},\theta^{\prime})
  = \frac{1}{2 \pi \sqrt{ \vert \mathbf{\Sigma} \vert}}
                     \exp{ \left(
                         - \frac{1}{2}
                           (\mathbf{x}-\mathbf{\mu})^T 
                           \mathbf{\Sigma}^{-1} 
                          (\mathbf{x}-\mathbf{\mu}) 
                           \right )   }  ,
\label{joint}
\end{equation}
where $\mathbf{x}= (\phi^{\prime},\theta^{\prime})$, $\vert \mathbf{\Sigma} \vert= 
\Sigma_{\phi^{\prime}\phi^{\prime}} \Sigma_{\theta^{\prime}\theta^{\prime}}-\Sigma_{\phi^{\prime}\theta^{\prime}}^2$ is the determinant 
of the covariance matrix $\mathbf{\Sigma}$ and 
$\mathbf{\mu}=(\mu_{\phi^{\prime}},\mu_{\theta^{\prime}})$
is the two-dimensional vector of 
the means of both parameter fluctuations.
Since we are addressing the joint distribution of fluctuations of both
parameters around their average pattern $\mathbf{\mu}\sim 0$.
In Fig.~\ref{fig05}a one sees the joint histogram of both parameter
fluctuations, while in Fig.~\ref{fig05}b the joint density function
in Eq.~(\ref{joint}) with the same covariance matrix $\mathbf{\Sigma}$ 
and means $\mathbf{\mu}$ is shown.
From the respective contour plots in Figs.~\ref{fig05}c and \ref{fig05}d 
one identifies a reasonable match. Moreover, one notices that both 
contour plots lean towards the left, indicating a negative correlation, 
i.e.~$\Sigma_{\phi^{\prime}\theta^{\prime}}<0$.

Being correlated, the parameter fluctuations must be
modelled as a pair of coupled variables. The model for the pair of
parameter fluctuations is described in the next section.

\section{The stochastic evolution of log-normal parameters}
\label{sec:evol}

For modelling the two fluctuations, $\phi^{\prime}$ and $\theta^{\prime}$,
we consider them as stochastic variables that evolve coupled to each
other according to a two-dimensional Markov process with two
independent stochastic Gaussian white noise sources.
The conditions that enable us to settle down such assumptions
will be addressed in the end.
Even in the case that the conditions do not hold rigorously,
it is possible to apply this framework, as discussed previously in
Ref.~\textcite{Paulo} and below.
For simplicity, in this section we are going to suppress the prime
symbol noticing that we only address the fluctuations. 

Our model is given by a system of two coupled Langevin equations, namely
\begin{eqnarray}
\label{eq2D}
\begin{bmatrix}
d \phi (t)\\
d \theta(t)
\end{bmatrix}
&=&
\begin{bmatrix}
{D_\phi}^{(1)} (\phi,\theta) \\
{D_\theta}^{(1)}(\phi,\theta)
\end{bmatrix}
dt \cr
 & & \cr
&+&
\begin{bmatrix}
g_{\phi \phi} (\phi,\theta) & g_{\phi \theta} (\phi,\theta)\\
g_{\theta \phi}(\phi,\theta)  & g_{\theta \theta} (\phi,\theta)
\end{bmatrix}
\begin{bmatrix}
dW^{(1)}_t \\
dW^{(2)}_t
\end{bmatrix}
\end{eqnarray}
where for either $X=\phi$ or $X=\theta$ 
\begin{widetext}%
\begin{equation}
D_{X}^{(1)}(\phi^{\ast},\theta^{\ast})= 
                 \lim_{\tau \rightarrow 0}\frac{1}{k!\tau} 
                 \left\langle
                 \left( X(t+\tau)-X(t) \right)
                 \right\rangle_{X(t)=\phi^{\ast},Y(t)=\theta^{\ast}}%
\label{D1_k}%
\end{equation}%
with $\langle\cdot\rangle$ meaning the average over all pairs
$(X(t),Y(t))$ of the parameter pair $(\phi,\theta)$ respectively and
where  $\mathbf{g}\mathbf{g}^{T}=\mathbf{D}^{(2)}$ with
\begin{equation}
\mathbf{D}^{(2)}=
\begin{bmatrix}
D^{(2)}_{\phi \phi} &  D^{(2)}_{\phi \theta} ,\\
D^{(2)}_{\theta \phi} &  D^{(2)}_{\theta \theta}  ,\\
\end{bmatrix} 
\end{equation}
such that, for the component $[XY]$ in matrix 
$\mathbf{D}^{(2)}$, one has
\begin{equation}
D_{XY}^{(2)}(\phi^{\ast},\theta^{\ast})=
                 \lim_{\tau \rightarrow 0}\frac{1}{k!\tau} 
                 \left\langle
                 \left( X(t+\tau)-X(t) \right)
                 \left( Y(t+\tau)-Y(t) \right)
                 \right\rangle_{X(t)=\phi^{\ast},Y(t)=\theta^{\ast}} \,.
\label{D2_ij}
\end{equation}%
\end{widetext}%
Here $\phi^{\ast}$ and $\theta^{\ast}$ label one specific bin.

The two stochastic contributions, $dW^{(1)}_t$ and $dW^{(2)}_t$, are two
independent Wiener processes, 
i.e.
\begin{subequations}
\begin{eqnarray}
\langle dW^{(n)}_t \rangle &=&  0 , \label{mean}\\
\langle dW^{(n)}_t dW^{(m)}_{t^{\prime}} \rangle &=& 
\delta_{mn}\delta(t-t^{\prime}) , \label{corr}
\end{eqnarray}
\label{twowiener}
\end{subequations}
with $n,m=1,2$.

The drift vector $({D_\phi}^{(1)}, {D_\theta}^{(1)})$ and the 
diffusion matrix $\mathbf{D}^{(2)}$ can be straightforwardly obtained 
from the sets of data using a recently package implemented
in $R$ and available as open source at CRAN platform\cite{Rinn2015}.  
\begin{table}[t]
\centering%
\begin{tabular}{|c|c|c|c|c|c|c|c|}
\hline
 & 1 & $\phi$ & $\theta$ & $\phi^2$ & $\phi\theta$ & $\theta^2$ & $R^2$\\
\hline
$D^{(1)}_\phi$  & -0.0085 & -0.7143 & 0.2812 & -- & -- & -- & 0.78 \\
\hline
$D^{(1)}_\theta$ & -0.0031 &  0.0293 & -0.5023 & -- & -- & -- & 0.67 \\
\hline
$g_{\phi\phi}$ & 0.2185 & 0.0918 & 0.2255 & 0.4850 & 0.2925 & 4.0541 & 0.93\\
\hline
$g_{\theta\theta}$ & 0.0360 & 0.0174 & -0.0128 & 0.0210 & 0.0245 & 1.5197& 0.92\\
\hline
$g_{\phi \theta}$ & -0.0111 & -0.0051 & -0.0158 & -0.0134 & 0.2936 & -0.1835 & 0.93\\
\hline
\end{tabular}
\vspace{0.5cm}
\caption{\protect
         Coefficients for the drift vector $\mathbf{D}^{(1)}$ and 
         matrix $\mathbf{g}$, obtained by the Langevin analysis.
         For each fit one also gives the $R^2$ value of the
         corresponding least square fit.}
\label{tab:coeff}
\end{table}

In Fig.~\ref{fig06} we plot the drift vector and the
diffusion matrix $\mathbf{D}^{(2)}$ obtained with this package.
The empirical results (bullets) of all drift and diffusion functions 
are compared with their polynomial fit (surfaces), linear forms for
the two drift coefficients and quadratic forms for the diffusion 
coefficients.
The derivation of matrix $\mathbf{g}$ from the diffusion matrix is described
in detail in Ref.~\textcite{vasconcelos}: since the diffusion matrix 
yields $\mathbf{D}^{(2)}=\mathbf{g}\mathbf{g}^T$ and has components fitted
by quadratic forms of $\phi$ and $\theta$, matrix $\mathbf{g}$ can also
have entries given by higher-order polynomials. 
Our results have shown that polynomials of degree one to the functions 
composing the drift vector and polynomials of degree two to the 
functions composing matrix $\mathbf{g}$ yield good results:
\begin{subequations}
\begin{eqnarray}
D^{(1)}(\phi,\theta) &\approx& a+b\phi+c\theta \, ,\label{fitD1}\\
g(\phi,\theta) &\approx& a+b\phi+c\theta+d\phi^2+e\phi\theta+f\theta^2 \, .\label{fitD2}
\end{eqnarray}
\label{fits}
\end{subequations}%
In Tab.~\ref{tab:coeff} we show the values of all polynomial coefficients
as well as the $R^2$ value of each fit.

The drift coefficients appear to depend linearly with respect to both 
parameter fluctuations:
the fluctuations are subjected to a restoring
force proportional to their amplitude.
The restoring force for $\phi$ (logarithmic mean) fluctuations is stronger 
than the one for $\theta$ (logarithmic variance) fluctuations, 
indicating a faster relaxation of the dynamics towards the equilibrium
value of the logarithmic mean, $\phi_0$, compared to the relaxation
of the logarithmic variance, which can be associated to a measure of
volume-price volatility.

In what concerns the diffusion matrix, one can detect a dominance of the 
quadratic term in $\theta$ for both diagonal terms in matrix $\mathbf{g}$. 
It seems that the variance of both fluctuations is governed mainly by the 
value of the logarithm variance itself. This makes sense since the 
variance of the fluctuations is responsible for the variance of the 
volume-price. 

\begin{figure*}[t]
\centering
\includegraphics[width=0.9\textwidth]{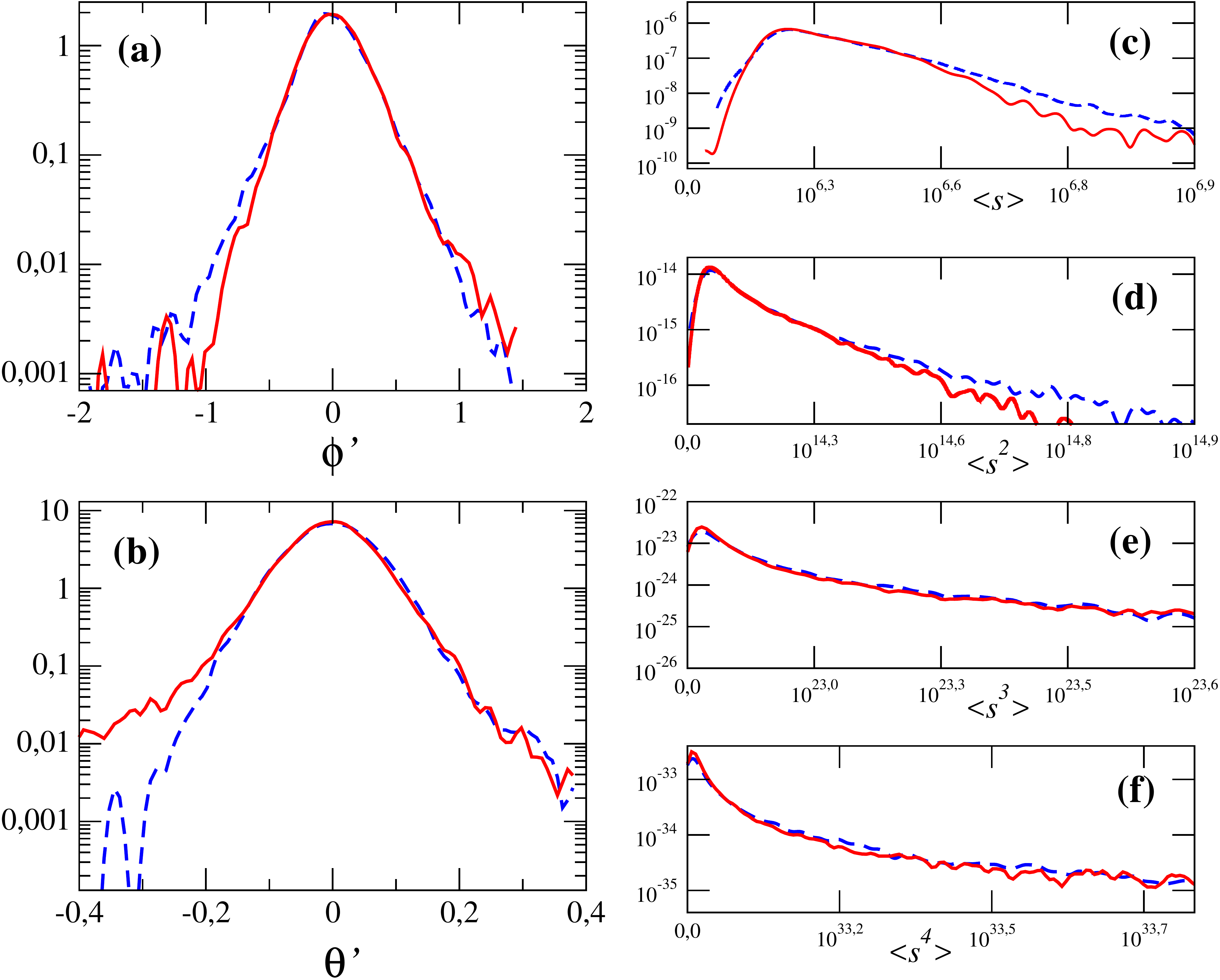}%
\caption{\protect
         \textbf{(a)} Comparison between the 
         PDFs of the empirical fluctuations of $\phi$ 
         (dashed line) and the modelled increments (solid line) in a 
         lin-log scale. 
         \textbf{(b)}  Same comparison is shown for $\theta$.
         \textbf{(c-f)} PDFs of the time series of the $n$-order moments
         $\langle s^n \rangle$ for $n=1,2,3,4$.
         The dashed line represents the empirical PDF and the solid 
         line is the PDF obtained from our model.
         Each series is composed by $16918$ points.}
\label{fig07}
\end{figure*}
\begin{figure*}[t]
\centering
\includegraphics[width=0.8\textwidth]{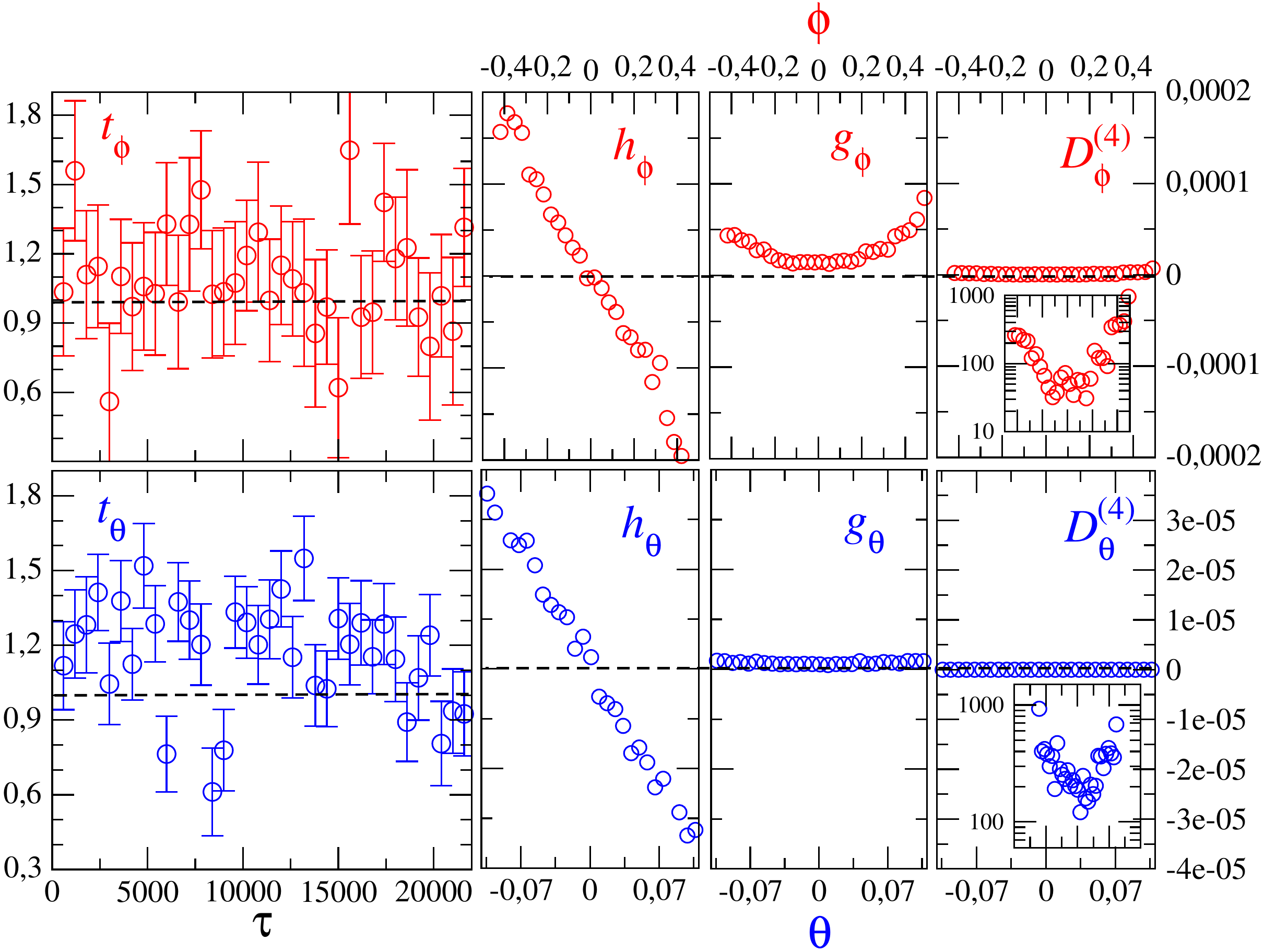}
\caption{\protect
         On the left, results of the Wilcoxon test to the detrended 
         time series, $\phi$ and $\theta$, showing the $t$-values.
         Here, both series are considered separately.
         In the middle, the drift $h$ and diffusion $g$, for both series.
         On the right, the fourth Kramers-Moyal coefficient $D^{(4)}$ is 
         plotted and in the inset the large values of $D^{(4)}/(D^{(2)})^2$ 
         plotted
         show that conditions for Pawula theorem are not exactly
         met (see text).}
\label{fig08}
\end{figure*}

\section{Approaching non-stationarity}
\label{sec:nonstationary}

In the two previous sections we introduce our framework to extract a
model for the quantities $\phi$ and $\theta$ parameterizing
the log-normal distribution fitting the volume-prices at each
10-minute frame.
As mentioned above,
we have assumed that all time dependency is incorporated in the 
two parameters and that they are decomposable into an
average pattern and fluctuations.
In this section we show that, with such a framework, 
one is able to fully characterize the non-stationary time series 
of the volume-price.

To that end, we first deduce the formula of all moments 
of the log-normal distribution in Eq.~(\ref{Log-normal_PDF}), 
for all integer $n$, namely
\begin{equation}
 \langle s^n \rangle = 
\int_{-\infty}^{+\infty} s^n p_{\phi,\theta}(s)ds =e^{n \phi + \frac{n^2 \theta ^2}{2}} \equiv F_n(\phi, \theta)\ .
\label{sn}
\end{equation}
It is a well known statistical resuls\cite{statisticsbook} that, if one has 
all the moments of a distribution, one can deduce its probability 
density function using a Fourier transform. 
Therefore, we need to derive from our models for the average parameter 
patterns and parameter fluctuations the evolution equation of
all moments as recently suggested in Ref.~\textcite{Paulo}.

Next, we will do this explicitly, by assuming that, similarly to 
$\phi$ and $\theta$, all moments are also separated in an average 
pattern and fluctuations around it.

The ``average'' $n$-order moments of the volume-price, 
$\langle \overline{s}^n \rangle$, is given
by the expression in Eq.~(\ref{sn}), substituting $\phi$ and $\theta$
by $\overline{\phi}$ and $\overline{\theta}$ respectively.

The fluctuations of the volume-price moments follow also the expression
in Eq.~(\ref{sn}) with the parameter fluctuations following the system of 
Langevin equations given in Eq.~(\ref{eq2D}) and can be derived
through the It\^o expansion\cite{vasconcelos} 
when differentiating the expression in Eq.~(\ref{sn}) with respect to time, 
yielding\cite{Paulo}
\begin{eqnarray}
d\langle s^n \rangle &=&  A_n(\phi(t),\theta(t))dt
     + B_n(\phi(t),\theta(t))dW_1 \cr
         & & \cr
     & & +C_n(\phi(t),\theta(t))dW_2 
\label{dsn}
\end{eqnarray}
with
\begin{subequations}
\begin{eqnarray} 
A_n(\phi(t),\theta(t)) &=& \frac{\partial F_n}{\partial \phi} h_1 + 
                                      \frac{\partial F_n}{\partial
                                        \theta} h_2  \cr
         & & \cr
&+& \frac{\partial^2
  F_n}{\partial\phi\partial\theta}(g_{11}g_{21}+g_{12}g_{22}) \cr
         & & \cr
&+& \frac{1}{2}\frac{\partial^2 F_n}{\partial\phi^2}(g^2_{11}+g^2_{12}) \cr
         & & \cr
&+& \frac{1}{2}\frac{\partial^2 F_n}{\partial\theta^2}(g^2_{21}+g^2_{22})  \, , \label{A}\\
         & & \cr
 B_n(\phi(t),\theta(t)) &=& \frac{\partial F_n}{\partial \phi} g_{11}+
                                        \frac{\partial F_n}{\partial
                                          \theta} g_{21} \, , \label{B}\\
         & & \cr
 C_n(\phi(t),\theta(t)) &=& \frac{\partial F_n}{\partial \phi} g_{12}+
                                         \frac{\partial F_n}{\partial
                                           \theta} g_{22} \, .\label{C}\\
                                       & & \cr
 \frac{\partial F_n}{\partial \phi} &=& n F_n (\phi, \theta)\\
  & & \cr
 \frac{\partial F_n}{\partial \theta} &=& n^2 \theta F_n (\phi, \theta)\\
  & & \cr
 \frac{\partial^2 F_n}{\partial \phi ^2} &=& n^2 F_n (\phi, \theta)\\
  & & \cr
 \frac{\partial^2 F_n}{\partial \theta ^2} &=& n^4 \theta ^2 F_n (\phi, \theta)\\
  & & \cr
 \frac{\partial^2 F_n}{\partial \phi  \partial \theta} &=& n^3 \theta F_n (\phi, \theta)\\
\end{eqnarray}
\label{ABC}
\end{subequations}
Equation (\ref{dsn}) is a non-homogeneous
stochastic differential equation with ``drift''
and ``diffusion''  functions depending on time.

In Fig.~\ref{fig07}a and \ref{fig07}b we compare the empirical
distribution of the fluctuations $\phi^{\ast}$ and $\theta^{\ast}$ 
respectively, with the corresponding modelled  distribution obtained
by integrating Eq.~(\ref{eq2D}). 
Within a range of fluctuations beyond one standard deviation
the modelled distribution fits well the empirical distributions.
The drift and diffusion functions were derived as described, adjusting
them afterwards to optimize the $R^2$ of the distributions in 
Figs.~\ref{fig07}(a-b) without deviating significantly from the least
square fits of the surfaces in Fig.~\ref{fig06}.


In Figs.~\ref{fig07}(c-f) we plotted the empirical and theoretical 
probability distributions of $\langle s^n \rangle, n=1,...,4$,
showing also good agreement. 
The empirical moments are obtained by replacing in Eq.~(\eqref{sn})
the original time series of $\phi$ and $\theta$.
Our model has a good fit in the first moments and can be used to model 
them since the theoretical and empirical distributions are very close 
to each other. Since the $\phi$ parameter is better modelled than the 
$\theta$ parameter, it is expected that for higher moments, when 
$\theta$ is dominant over $\phi$ we do not achieve such good results. 

In order to apply a Langevin model, it is necessary 
that the fluctuations time series, $\phi^\prime$ and $\theta^\prime$, are 
Markovian. 
To test the Markov property of the data series, we compute the transition 
probabilities $p(x_{1},\tau_1|x_{2},\tau_2;x_3,\tau_3)$ and 
compare it with the two point conditional probabilities 
$p(x_{1},\tau_1|x_{2},\tau_2)$. 
For that we use Wilcoxon rank-sum test \cite{wilcoxon}.
Values of $t/t_0$ close to 1, indicates the data has the Markovian 
property, which, as shown in Fig.~\ref{fig08} (left), is observed
already for the smallest values of time increments.

We also computed the $D_4$ function for both time fluctuations series,
to test if the conditions to apply Pawula Theorem are fulfilled\cite{risken}.  
The results obtained are in Figure \ref{fig08} (right). Here, we can see 
that $D_4$ is almost zero for both time series, though the fourth Kramers-Moyal
coefficient is not negligible in front of diffusion. Related to this
condition that is not rigorously fullfilled may be the deviations
obtained before optimizing the best set of coefficients that fit
the marginal distributions in Fig.~\ref{fig07}.

\section{Discussion and Conclusions}
\label{sec:conclusions}

The main goal of this paper was to model the non-stationary time series of the volume-price. By assuming that the log-normal had the best fit to the data in each 10-minutes window, this goal resumes to the one of studying the parameters $\phi$ and $\theta$ of this distribution, which are themselves stochastic variables. We were able to show that we can describe the time series of these parameters by decomposing the variables as a sum of two terms: one accounting for the daily pattern and another regarding the fluctuation around that average pattern. The fluctuations are modelled using a system of Langevin equations whose coefficients we retrieved from our empirical data. From here, we proposed a framework to reconstruct the evolution of all the moments of the volume-price distribution. 

This work leaves some open questions to be answered. It is true that we achieved a good model to the $\phi $ fluctuations, but we could not match this result to the $\theta$ fluctuations. One possible explanation is related with the outliers: we removed all the points which did not lie in a $5 \sigma$ interval from the mean. However, when we plotted the time series without the outliers, there were still some extreme values that look more like measurement errors than fluctuations. We chose to use the $5 \sigma$ criterion because we tried to minimize the number of points taken from our sample in order to let our data as close as possible to the original one. However, if one prefers to choose a stricter criteria, like using a $3\sigma$ interval, then the time series would have lesser outliers and maybe the results would be more easily modelled.
Some optimization adjustments were also needed when reproducing the distribution
of the values of the four distribution moments varying in time.
Probably associated to these deviations is the non-zero fourth Kramers-Moyal coefficient, which indicates possible deviation when assuming the Fokker-Planck truncation.

There are many models in the literature that enable us to study and to model stochastic time series such as autoregressive models\cite{statistics2}, moving-average models and autoregressive integrated moving average models \cite{timeseries,numericalrecipes}. One may ask, why did we choose the Langevin model instead of all the others. One strong argument in favor of this model is that it not only allows us to describe the evolution of our time series, but it may also give us an equation, Fokker-Planck equation, to describe the evolution of the volume-price distribution. Further work should be done in trying to extract such an equation from the equations we already have. If one is able to do this, then we would have much more information about the volume-price evolution and we could apply this information to the computation of the \textit{Value at Risk} or other risk measures\cite{riskbook}.
A possible open issue related to this point is the modelling of data based in
stochastic partial differential equations.

A comparison between our model and the classic models that have been used for studying time series would be an interesting work to develop in the future. It is true that our model has the advantage of being able to produce an equation to the evolution of the distribution of the volume price. The results achieved with our model may be better than the ones achieved by the classical models. In order to test this hypothesis, we should do this comparative study.

Finally, this work gave us important insight in the study of non-stationary time series and we have proposed here a methodology that we beliebe is useful in numerous fields. This framework is general enough to be applied to other markets besides the NYSE and also to other fields of study like physiology, when we are trying to study the heart interbeat intervals or geology, in order to study seismic time series. 

\section*{Acknowledgments}

JE thanks Philipp Maass for the opportunity of finishing her master’s 
dissertation at the University of Osnabr\"uck and also the
Program Erasmus sponsored by the European Union. 

\section*{References}

\bibliographystyle{apsrev4-1}
\bibliography{bibliography}

\end{document}